\documentclass[runningheads,a4paper]{llncs}

\usepackage{amssymb}
\usepackage{graphicx}

\usepackage{url}
\urldef{\mailf}\path|felix@neuro.uni-bremen.de|    
\urldef{\mailp}\path|pawelzik@neuro.uni-bremen.de|    
\newcommand{\keywords}[1]{\par\addvspace\baselineskip
\noindent\keywordname\enspace\ignorespaces#1}
\newcommand{\figref}[1]{{Fig.~\ref{#1}}}
\newcommand{\subfigref}[2]{{Fig.~\ref{#1}~#2}}

\newcommand{\longforref}[1]{{Eq.~(\ref{#1})}}

\newcommand{\appropto}{\mathrel{\vcenter{
  \offinterlineskip\halign{\hfil$##$\cr
    \propto\cr\noalign{\kern2pt}\sim\cr\noalign{\kern-2pt}}}}}

\begin{document}

\mainmatter  

\title{Bubbles, Jumps, and Scaling from Properly Anticipated Prices}

\titlerunning{Bubbles, Jumps, and Scaling from Properly Anticipated Prices}

%
%
\author{Felix Patzelt\thanks{Currently supported by the Volkswagen-Foundation}, Klaus Pawelzik}
\institute{Uni Bremen\\
\mailf\\
\mailp\\
\url{http://neuro.uni-bremen.de}}

%
%

\tocauthor{Authors' Instructions}
\maketitle

\begin{abstract}
Prices in financial markets exhibit extreme jumps far more often than can be accounted for by external news. Further, magnitudes of price changes are correlated over long times. These so called stylized facts are quantified by scaling laws similar to, for example, turbulent fluids. They are believed to reflect the complex interactions of heterogenous agents which give rise to irrational herding. Therefore, the stylized facts have been argued to provide evidence against the efficient market hypothesis which states that prices rapidly reflect available information and therefore are described by a martingale. Here we show, that in very simple bidding processes efficiency is not opposed to, but causative to scaling properties observed in real markets. Thereby, we link the stylized facts not only to price efficiency, but also to the economic theory of rational bubbles. We then demonstrate effects predicted from our normative model in the dynamics of groups of real human subjects playing a modified minority game. An extended version of the latter can be played online at seesaw.neuro.uni-bremen.de.
\keywords{Bubbles, Stylized Facts, Efficient Markets, Minority Games}
\end{abstract}

\section{Introduction}

Many studies of financial datasets emphasise scaling properties \cite{farmer99scaling}.  Further, large jumps in price time series often cannot be attributed to external events \cite{joulin2008news}. Some claim that these findings contradict the efficient market hypothesis (EMH) \cite{lux1999multi-agent}. However, the EMH foremost states, that no systematic profit is possible
from observing previous prices $p$ because predictable price changes are
eliminated by traders exploiting them \cite{samuelson_prices}. This property
holds well as a stylized concept \cite{Fama1998EfficiencyReturnsBehaviour}.
The distribution of log-returns $r(t) = \log(p_t / p_{t-1})$ would be Gaussian only if further assumptions like a fast convergence according to the central limit theorem would hold true which is not neccessarily the case. Furthermore, there are also economic models with perfectly rational traders that can exhibit ``rational bubbles'' where prices deviate far from fundamental values \cite{brunnermeier08}. Hence, bubbles and crashes do not disprove the EMH.

Here, we establish a systematic link between these endogenous mechanisms for bubbles and crashes on the one hand and the stylized facts on the other hand. We show, that price efficiency in a simplistic bidding
process links bubbles with the most prominent stylized facts of financial price time series. While the model is rather abstract and currently tied to a specific pricing rule, many qualitative and quantitative features of real returns are captured both numerically and analytically.
Our model further makes directly testable predictions, some of which were confirmed in behavioral group experiments.

\section{The Model}

Consider a market with $N$ agents: $N_s$ speculators and $N_r$ random traders. At discrete times $t$ each agent places a market order to either buy or sell one unit of an asset (e.g. a stock). Thereby, agents contribute to either the demand $d_t$ or to the supply $s_t = N - d$. For simplicity we only allow market orders, that is, $d_t$ and $s_t$ do not depend on the price at which the orders are executed. Note, that the latter is generally not known a priori at stock-, foreign exchange-, and similar markets. We further require: 1. Increasing $d_t$ increases the price while increasing $s_t$ decreases the price. 2. The price $p_t$ is invariant to the traded volume: scaling $d_t$ and $s_t$ by the same factor yields the same price. This allows e.g. for some  not executed orders as long as the same fraction of buy and sell orders are affected. Therefore,
\begin{equation}
	p_t = \frac{d_t}{s_t} = \frac{d_t}{N - d_t}
	\label{for:pricing}
\end{equation}
which naturally possesses the correct unit. 

Agents make their decisions stochastically and we postulate price efficiency: the probability for a speculator to buy at each time $t$ is chosen such that the expectation value of the new price given all previous observations
\begin{equation}
	\langle p_t\, |\, p_{t-1}, p_{t-2}, \dots \rangle \stackrel{!}{=} p_{t-1}
	\label{for:martingale_price}
\end{equation}
is the same as the previous price. This condition may be violated only if $d_t > N_s + N_r / 2$ or if $d_t < N_r / 2$. In these cases, it is impossible to be price efficient in this model due to the discretization of the traded assets. However, for $N_s \gg N_r$ and $N_s \gg 1$, we consider this boundary effect acceptable. 

\begin{figure}
	\includegraphics{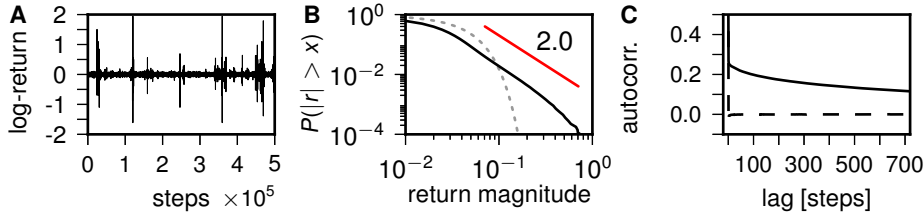}
	\caption{Price efficient model with $N_s = 10000$ speculators and $N_r = 10$ random (coin flipping) traders. A: Time series. B: Complementary cumulative distribution of log return magnitudes (solid black line) and a Gaussian with the same variance (dashed gray line). Straight line: analytical result. C: Autocorrelation of the log returns (dashed line) and of their magnitudes (solid line).}
	\label{fig:series}
\end{figure}

A model time series is shown in \subfigref{fig:series}{A}. The distribution of log-returns is power-law tailed. The exponent in the cumulative distribution approaches $\xi = 2$ for large systems (\subfigref{fig:series}{B}). Finite size effects or large $N_r / N_s$ increase $\xi$ (not shown). Log-returns are uncorrelated while their magnitudes are correlated for long periods of time (\subfigref{fig:series}{C}) reflecting realistic volatility clustering.

\section{Analytical Results}

To obtain an explicit solution to \longforref{for:martingale_price}, a good approximation for large $N$ is to require efficient demands instead of efficient prices:
\begin{equation}
	\langle d_t, |\, d_{t-1}, d_{t-2}, \dots \rangle \stackrel{!}{=} d_{t-1}.
	\label{for:martingale_demand}
\end{equation}
Since agents choose stochastically, the demands generated by the speculators and random traders each are binomially distributed. \longforref{for:martingale_demand} is fulfilled, if the probability for each speculator to buy at time $t$ is
\begin{equation}
	P(\textrm{buy}\, |\, d_{t-1}) = \frac{1}{2} + \frac{d_t - N / 2}{N_s}
	\label{for:p_buy_effdemand}
\end{equation}

\subfigref{fig:analytics}{A} shows a comparison of \longforref{for:p_buy_effdemand} with a numerical optimization with respect to \longforref{for:martingale_price}. For efficient prices, there is a slight drift away from the system boundaries that is not present for efficient demands. However, this difference decreases with an increased system size $N$.

\longforref{for:p_buy_effdemand} further shows why a small number of random agents is important for a finite system. For $N_r = 0$, we obtain $P(\textrm{buy}\, |\, d_{t-1}) = d_{t-1} / N$. Then, if by chance the system ends up in the boundary states $d = 0$ and $d = N$, it can never leave unless we allow for a violation of \longforref{for:martingale_price} as discussed above. An alternative to random agents would be a reset rule like $P(\textrm{buy}\, |\, d_{t-1} \in \{0, N\}) = 1/2$.

\subsection{Stationary solution}

\begin{figure}
	\centering
	\includegraphics{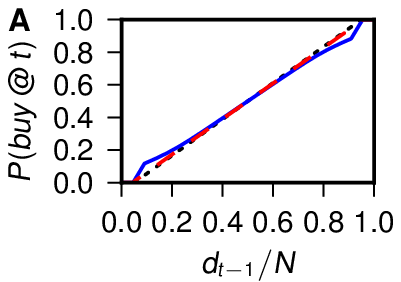}
	\hspace{2cm}
	\includegraphics{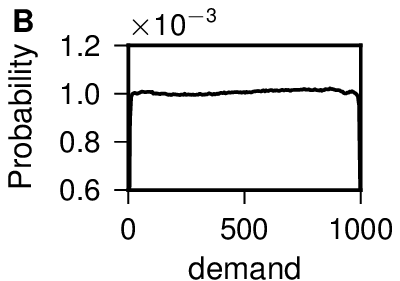}
	\caption{A: Probability for an agent to buy at time $t$ for different previous demands $d_{t-1}$ normalized by the system size $N = N_s + N_r$. Fraction of random traders: $N_r / N_s = 0.1$. The demand efficient solution is given by \longforref{for:p_buy_effdemand} (dotted black). The price efficient solutions for $N = 22$ (solid blue) and $N = 110$ (dashed red) were obtained by numerical optimization.
	B: Distribution of demands in a simulation of $N_s = 1000$ speculators and $N_r = 1$ random trader for $10^8$ time steps. }
	\label{fig:analytics}
\end{figure}

For large $N_s$, we can neglect the random agents, and the difference between price- and demand efficiency. We thus consider $N$ agents who buy at each time $t$ with probability $d_{t-1} / N$. The stationary demand distribution $\pi$ then satisfies
\begin{equation}
	\pi_j = \sum_{i=0}^{N} \pi_i\, \pi_{ij}, \quad\quad \pi_{i,j} = {N \choose j}\ \left(\frac{i}{N}\right)^{j}\ \left(1 - \frac{i}{N}\right)^{N -j}
	\label{for:stationary_solution}
\end{equation}
where the probability to move from state $i = d_{t-1}$ to state $j = d_t$ is given by the transition matrix $\pi_{ij}$.
For large $N$, \longforref{for:stationary_solution} is satisfied by the uniform distribution. To show this, we first divide by $\pi_i = \pi_j = \textrm{const}$, and obtain
\begin{equation}
	1 = \sum_{x=0}^{1}\ {N \choose j}\  x^{j}\ (1 - x)^{N -j},\quad \textrm{with}\quad x = \frac{i}{N}.
	\label{for:stationary_solution_pt2}
\end{equation}
For large $N$, we can replace the sum over $x \ll 1$ with an integral. The right hand side of \longforref{for:stationary_solution_pt2} then reads
\begin{eqnarray}
	{N \choose j}\ N \int_0^1 x^j (1 - x)^{N - j} dx &=& {N \choose j}\ N\ \frac{\Gamma(j+1)\Gamma(N-j+1)}{\Gamma(N+2)}\\
	&=& \frac{N}{N+1} \stackrel{N \gg 1}{\longrightarrow} 1 \hspace{2.5cm}\Box
\end{eqnarray}
\subfigref{fig:analytics}{B} shows the demand distribution for a simulation of the price efficient model. It is uniform except for the very edges where it drops sharply. For higher ratios $N_r / N_s$, the edges can also exhibit peaks.

\subsection{Tail Exponent}

The log return for two subsequent demands $d$ and $d'$ can be expressed as
\begin{equation}
	r = \log\left( \frac{d'}{N-d'}\frac{N-d}{d}\right) \approx \Delta \left(\frac{1}{d} - \frac{1}{N-d}\right), \quad \textrm{where}\quad \Delta = d' - d.
\end{equation}
The approximation is obtained by expanding for small $\Delta$ up to the first order. This is possible, because the standard deviation for the binomial distribution is proportional to $\sqrt{N}$ and 
vanishes for demands close to zero or close to $N$. Hence, the distribution of $\Delta$ will be very localized for large $N$. Fluctuations in $r$ are then dominated by $d$, especially for $d \ll N$, and $N-d \ll N$. Due to the symmetry with respect to $N/2$, we now analyze only the case $d \ll N$ where $r \approx \frac{\Delta}{d}$. For the scaling of the tail of the return magnitudes, the shape of $p(\Delta)$ is negligible. The expected fluctuations in $r$ can be expressed by
\begin{equation}
	\langle r^2 | d \rangle \approx \left( \frac{\langle \Delta \rangle}{d} \right)^2 \approx \frac{1}{d} := \tilde r^2.
\end{equation}
Using the probability integral transform, and $p(d) = \textrm{const.}$ yields
\begin{eqnarray}
	p(\tilde r) \propto |\tilde r|^{-3}, \quad \textrm{and therefore} \quad p(r) \appropto |r|^{-3}
\end{eqnarray}
for sufficiently large $N$ and $r$, and in agreement with simulations (\subfigref{fig:series}{B}).

\section{Experiments}

Testing for tail exponents or stationary distributions with limited time and subjects appears impossible. However, the uniform distribution of demands implies a dynamics which spends significant amounts of time near the system boundaries, that is, in bubble phases. This testable prediction reflects that a mean reverting trend may be easily exploited and eliminated by traders. 
We let subjects play a game (an extension is playable online \cite{seesawgame}) where players $i$ in each round chose $c_{i,t} = 1$ or $c_{i,t} = -1$ before a countdown ran out. These choices correspond to the market orders in the model above. No decision was registered as $c_{i,t} = 0$. A superplayer chose $C_t = -\sum_i c_{i, t-1} = -d_{t-1}$. The players whose choices were in the minority won $1$ point. Due to the superplayer, betting against the change in the other players decisions is rewarded. Note, that each new round is a nash equilibrium: If all players repeat their choices, the outcome $d_{t} - d_{t-1}$ (a linearized return) is zero. A single player who changes, loses. Yet, players did not stay in these equilibria.

\begin{figure}
	\includegraphics{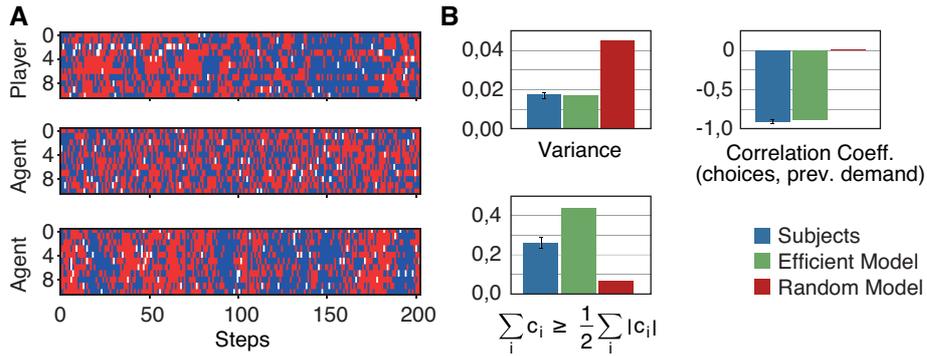}
	\caption{A: Top: choices in an experiment with 11 Subjects. Red and blue correspond to the choices $c_i = \pm 1$, white to skipping a round. Middle: model with equal probabilities for $c_i \pm 1$. Bottom: demand efficient model. Skipping probabilities for the models were equal to the experiments.  B: Variance of the outcomes, correlation coefficient of the player choices with the respective last excess demand, and the probability for a bubble. The latter is quantified by the relative number of rounds, where one choice was made by twice as many subjects (agents) as the other choice.}
	\label{fig:experiments}
\end{figure}

\figref{fig:experiments} shows the subjects' choices and two models: agents chosing by flipping a coin, and agents betting demand efficient on average. The experiment and the efficient model show clusters where one choice is preferred. These bubble phases are absent for coin flipping agents. Efficient betting causes a decrease in the outcome variance, but increases the probability for a bubble phase. This effect is significant, but not  as strong for the real subjects as for the efficient model. This may be due to a heterogenity of players or the use of other information which is not captured by the simple model. Nevertheless, players bet against the superplayer and therefore against mean reversion as much as the efficient model.

\section{Summary}

We presented an analytically tractable model which relates price efficiency to bubbles, power-law log returns and volatility clusters. The lack of mean reversion leads to a uniform demand distribution. The non-linear price causes the system to be more susceptible in bubble phases. This is analogous to, for example, many buyers betting up the price of a scarce resource. Then, in absolute terms small fluctuations in the availability of said resource may lead to large relative price changes. Another analogy would be a liquidity crisis.

We successfully tested model predictions with human subjects. Instead of a player payoff based on the excess demand like in other minority games, we use the return. This correctly compares the price at which an asset is sold by an agent not to a fundamental price, but to the price at which the agent bought said asset, and vice versa. Our game combines information efficiency as in minority games with bubbles as in majority games in a simpler way than the \$-game \cite{andersen2003} \cite{giardina2003}, and without the necessity to fine tune to a phase transition (for an overview of games, see \cite{challet2004market_chapter}). Even if player choices were not efficient, adjusting their impact based on our payoff rule (trading success) is a learning algorithm allowing for collective efficiency with respect to the information available to the agents. \cite{patzelt2013instability}.

\bibliography{references}

\end{document}